\documentclass[12pt]{amsart}
\usepackage[english]{babel}
\usepackage{amsmath}
\usepackage{amsfonts,amsthm}
\usepackage{amssymb}

\newtheorem{lemma}{Lemma}
\newtheorem{prop}[lemma]{Proposition}
\newtheorem{teo}[lemma]{Theorem}
\newtheorem{exam}[lemma]{Example}
\newtheorem{coral}[lemma]{Corollary}

\begin{document}

\title[Anisotropic integral operator in superconductivity]{An
anisotropic integral operator in high
temperature superconductivity}

\author{Boris Mityagin}

\address{Department of Mathematics,
The Ohio State University,
 231 West 18th Ave,
Columbus, OH 43210, USA}

\email{mityagin.1@osu.edu}



\dedicatory{Dedicated to Plamen Diakov on the occasion of his
sixtieth birthday}

\begin{abstract}
A simplified model in superconductivity theory studied by P. Krotkov
and A.~Chubukov \cite{KC1,KC2} led to an integral operator $K$~---
see (1), (2). They guessed that the equation $E_0(a,T)=1$ where
$E_0$ is the largest eigenvalue of the operator $K$ has a solution
\[
 T(a)=1-\tau(a) \text{ with } \tau (a) \sim a^{2/5} \leqno{(*)}
\]
when $a$ goes to $0$. $\tau(a)$ imitates the shift of critical
(instability) temperature.

 We give a rigorous analysis of an anisotropic integral operator
$K$ and prove the asymptotic ($*$)~--- see Theorem 8 and Proposition
10.

Additive Uncertainty Principle (of Landau-Pollack-Slepian [SP],
\cite{LP1,LP2}) plays important role in this analysis.
\end{abstract}

\maketitle

0. Many models of high temperature superconductivity \cite{LV} lead
to the family of integral operators with anisotropic kernels.
Structure and spectral analysis of these operators could be
difficult and quite interesting because standard analytical methods
(perturbation theory, Fourier transform, etc.) do not necessarily
help us.

P.~Krotkov and A.~Chubukov \cite{KC1,KC2} [see \cite{KC2}, section
B.2, (46)--(60)] simplified one of local Eliashberg gap equations by
dropping the Matsubara frequency summation and came to the operator
in $L^2(\mathbb{R})$
\begin{equation}\label{M-1}
    K\equiv K_{aT}:\, f(x)\to \int_{-\infty}^\infty K(x,y)f(y)\,dy,
\end{equation}
where
\begin{equation}\label{M-2}
    K(x,y)=\dfrac1\pi\cdot \dfrac1{T^2+(x-y)^2+a^2(x^2+y^2)^2}.
\end{equation}
If $a=0$ the operator is a convolution
\begin{equation}\label{M-3}
    k*f\quad
    \text{where }k(x)=\dfrac1\pi\cdot\dfrac1{T^2+x^2}
\end{equation}
and its Fourier transform $\widetilde{k}(s)=\dfrac1Te^{-T|s|}$.
Therefore, $\|K_{0T}\|=1/T$,
\begin{equation}\label{M-4}
    \|K_{01}\|=1
\end{equation}
and its spectrum $\sigma(K_{01})=[0,1]$ or $\sigma(K_{0T})=[0,1/T]$.
If $a>0$, $K(x,y)=K(y,x)>0$,
\begin{equation}\label{M-5}
    \iint K^2\,dxdy<\infty
\end{equation}
so $K$ is a self-adjoint Hilbert--Schmidt operator, its spectrum
$\{E_n(a)\}$ is discrete and by monotonicity
\begin{equation}\label{M-6}
    \|K_a1\|<1.
\end{equation}
Subspaces $H^e$ and $H^o$ in $L^2(\mathbb{R})$ of even and odd
functions are invariant so we'll consider restrictions
\begin{equation}\label{M-7}
    K^e=K|H^e
    \quad
    \text{and}
    \quad
    K^o=K|H^o
\end{equation}
and their spectra. Let $E^e(T,a)$ and $E^o(T,a)$, $a>0$ be the
largest eigenvalues of $K^e$ and $K^o$, correspondingly. By
\eqref{M-3} and \eqref{M-6}
\begin{equation}\label{M-8}
    E^e(a,1)=1-\varphi(a),\,\varphi>0,
\end{equation}
\begin{equation}\label{M-9}
    E^o(a,1)=1-\psi(a),\,\psi>0,
\end{equation}
and $E^o(a,T)=1$ for some
\begin{equation}\label{T-10}
    T=1-\tau(a),\,\tau>0.
\end{equation}

The toy model in \cite{KC2}, Sect B.2, views $\tau(a)$ as an
imitation of the shift of critical (instability) temperature where
$a$ is the dimensionless quantity proportional to both the curvature
and interaction (see (36), (37) in [4] for details). Heuristic
manipulations (46)--(62) in \cite{KC2} intended to make us believe
that $\tau(a)\sim c\,a^{2/5}$. Maybe, this is quite remarkable that
the potential ``$2/5$'' appears in this analysis. We will show in
this essay that indeed for $a>0$ small enough
\[
 c_1a^{2/5}\leq\tau(a)\leq c_2a^{2/5}
\]
where $c_1,c_2>0$ are absolute constants (see Prop.~\ref{prop-10} in
Sect 4 below).

1. But first, we will find good estimates of the shift $\psi$ in
\eqref{M-9}, i.e., the behavior of the largest eigenvalue of
$K^o\in\eqref{M-7}$. Of course,
\begin{equation}\label{M-11}
   E(a)=\|K_a\|;\, E^e(a)=\|K_a|H^e\|;\,
    E^o=\|K_a|H^o\|
\end{equation}
because $K$ is a self-adjoint compact operator and $H^e$, $H^o$ are
its invariant subspaces.
\begin{lemma}\label{lem-M-1}
For $a>0$
\begin{equation}\label{M-12}
    1>E(a)=E^e(a)>E^o(a)
\end{equation}
and
\begin{equation}\label{M-13}
    E(0)=E^e(0)=E^o(0)=1.
\end{equation}
\end{lemma}
\begin{proof}
As \eqref{M-3} shows, after Fourier transform
\begin{equation}\label{M-14}
    \widetilde{K}_{01}=\mathcal F^{-1} K_{01}\mathcal F
\end{equation}
is a multiplier-operator
\begin{equation}\label{M-15}
    \widetilde{k}:\,
    \varphi(s)\to e^{-|s|}\varphi(s)
\end{equation}
and
\begin{equation}\label{M-16}
    \langle\widetilde{k}\varphi,\varphi\rangle=
    \int_{-\infty}^\infty e^{-|s|}|\varphi(s)|^2\,ds,
\end{equation}
\eqref{M-13} follows easily. But $\widetilde{K}_{01}$ is not compact
and the norms in \eqref{M-13} are not attained as values of a
quadratic form
\begin{equation}\label{M-17}
    \langle Kf,f\rangle,
    \quad\|f\|=1,\,f\in H^e\,\,\text{or}\,\, H^o.
\end{equation}
{Or if you wish they are ``attained'' if $f^e=\delta(x)$ and
$f^o=\delta^\prime(x)$}.

If $a>0$, the norm is attained as, say,
\begin{equation}\label{M-18}
    E^o(a)=\langle K_ag,g\rangle,\,\,
    \exists\,g\in H^o,\,\, \|g\|=1
\end{equation}
where $K_ag=E^o(a)g$. But then
\begin{equation}\label{M-19}
    g_*(x)=|g(x)|\quad
    \text{is even, }\|g_*\|=\|g\|=1,
\end{equation}
and
\begin{equation}\label{M-20}
    E^e(a)\geq \langle Kg_*,g_*\rangle> \langle Kg,g\rangle=E^o(a).
\end{equation}
We have a strict inequality in \eqref{M-20} because $K(x,y)>0$ and
an odd $g(x)\not=0$ is negative and positive on some subsets of
positive measure. So
\begin{equation}\label{M-21}
 \iint\limits_{\mathbb{R}^2} K(x,y)g(x)g(y)\,dxdy <
  \int\limits_{\mathbb{R}^2} K(x,y)|g(x)|\cdot|g(y)|\,dxdy.
\end{equation}
Indeed, if $g$ is odd and not identically zero put
\[
 G^+=\{x\in\mathbb{R}^1|\,g(x)>0\},
\]
\[
 G^-=\{x\in\mathbb{R}^1|\,g(x)\leq0\}.
\]
Then measures $\lambda(G^+)>0$, $\lambda(G^-)>0$ are positive and
\[
 \iint_{\mathbb{R}^2}K(x,y)g(x)g(y)\,dxdy
\]
= sum of four integrals $\displaystyle\iint_{G^\pm\times G^\pm}$.
Two of then (over $G^+\times G^+$ and $G^-\times G^-$) are positive
and two (over $G^+\times G^-$ and $G^-\times G^+$) are negative
because $K(x,y)>0$ everywhere with the excess being equal to
\[
 -\left(\int_{G^+\times G^-} Kg(x)g(y)\,dxdy\right)>0.
\]

Almost the same argument shows that $E(a)<1$. Indeed,
\begin{equation}\label{M-22}
    K_a(x,y)<K_0(x,y)
\end{equation}
everywhere on $\mathbb{R}^2$. So
\begin{equation}\label{M-23}
 E(a)=\langle K_ag^*,g^*\rangle
\end{equation}
for some $g^*\in L^2(\mathbb{R}^1)$, $\|g^*\|=1$, $g^*\geq0$, and
\eqref{M-22} implies
\begin{equation}\label{M-24}
  E(a)<\langle K_0g^*,g^*\rangle\leq\|K_0\|=1.
\end{equation}
Lemma~\ref{lem-M-1} is proven.
\end{proof}

2. From now on we analyze the integral operator $K=K_a$, $a>0$, with
a kernel \eqref{M-2}, $T=1$. By \eqref{M-7} we can consider it in
the block-form
\begin{equation}\label{M-25}
    K\sim\begin{bmatrix}
     K^e & 0\\
     0 & K^o
    \end{bmatrix}
\end{equation}
where $K^e$, $K^o$ are integral operators but their kernels are not
uniquely determined because, say for $h\in H^o$ and $A(x,y)=A(x,-y)$
\begin{equation}\label{M-26}
    \int A(x,y)h(y)\,dy=0
\end{equation}
any way. To analyze $K^o$ we change a kernel \eqref{M-2} to its
antisymmetrization
\begin{equation}\label{M-27}
    K^\prime(x,y)=\dfrac12[K(x,y)-K(x,-y)\
\end{equation}
and still have a representation
\begin{equation}\label{M-28}
    K^of=\int K^\prime(x,y)f(y)\,dy,\,\, \forall f\in H^o.
\end{equation}
Of course,
\begin{equation}\label{M-29}
    \int K^\prime(x,y)g(y)\,dy=0,\,\, \forall g\in H^e,
\end{equation}
this is a twin of \eqref{M-26}.

Explicit formula for $K^\prime$ is
\begin{multline}\label{M-30}
    K^\prime(x,y)=\dfrac1\pi
     \left[\dfrac1{1+(x-y)^2+a^2(x^2+y^2)^2}-\right.\\
    \left.-\dfrac1{1+(x+y)^2+a^2(x^2+y^2)^2}\right]=\\
    =\dfrac1{2\pi}\cdot \dfrac{4xy}
    {(1+x^2+y^2+a^2(x^2+y^2)^2)^2-4x^2y^2},
\end{multline}
and in polar coordinates $(r,\varphi)$ with
\begin{equation}\label{M-31}
    x=r\cos\varphi,\,
    y=r\sin\varphi,\,\,
    0<r<\infty,\,0\leq\varphi<2\pi
\end{equation}
\begin{equation}\label{M-32}
    K^\prime(x,y)=\dfrac1\pi\cdot
    \dfrac{r^2\sin2\varphi}{(1+r^2+a^2r^4)^2-r^4(\sin2\varphi)^2}.
\end{equation}

We want to get estimate for $E^o(a)$ from below by choosing a test
function
\begin{equation}\label{M-33}
    f_*(x)=x\exp\left(-\dfrac12h^2x^2\right),
\end{equation}
$h>0$ to be specified later, $hH=1$, and doing explicit calculations
of quadratic forms. So
\begin{equation}\label{M-34}
    E^o(a)\geq \dfrac{\langle K^of_*,f_*\rangle}{\langle
    f_*,f_*\rangle}=1-\alpha(a).
\end{equation}
\begin{prop}\label{prop-M-2}
For a kernel \eqref{M-30}, \eqref{M-32} with notations \eqref{M-33},
\eqref{M-34} for small enough $a>0$
\begin{equation}\label{M-35}
    \alpha(a)\leq Ca^{2/5},
\end{equation}
$C$ an absolute constant, $C<3$.
\end{prop}
\begin{proof}
\begin{multline}\label{M-36}
    \langle f_*,f_*\rangle=\int_{-\infty}^\infty x^2
    \exp(-h^2x^2)\,dx= H^3\int_{-\infty}^\infty
    \int_{-\infty}^\infty (hx)^2\exp(-(hx)^2)\,d(hx)=\displaybreak[2]\\
    =2H^3\int_0^\infty y^2e^{-y^2}\,dy=
    H^3\int_0^\infty u^{1/2}e^{-u}\,du=
    H^3\Gamma(3/2)=\dfrac{\sqrt{\pi}}2H^3.
\end{multline}
Next, by \eqref{M-31}, \eqref{M-32},
\begin{equation}\label{M-37}
    f_*(x)f_*(y)=xy\exp\left(-\dfrac12h^2(x^2+y^2)\right)=
    \dfrac12r^2\exp\left(-\dfrac12h^2r^2\right)\sin2\varphi,
\end{equation}
and
\begin{multline}\label{M-38}
    \langle K^of_*,f_*\rangle=\langle K^\prime f_*,f_*\rangle=
    \iint_{\mathbb{R}^2} K^\prime(x,y)f_*(x)f_*(y)\,dxdy=\displaybreak[2]\\
    =\dfrac1\pi\int_0^\infty \int_0^{2\pi}
    \dfrac{e^{-\frac12h^2r^2}r\frac12r^4(\sin2\varphi)^2\,d\varphi
    dr}{C^2-A^2(\sin2\varphi)^2}
\end{multline}
where
\begin{equation}\label{M-39}
    C=1+r^2+a^2r^4,
\end{equation}
\begin{equation}\label{M-40}
    A=r^2;\, C^2=A^2+B^2,
\end{equation}
and
\begin{equation}\label{M-41}
    B=\sqrt{C^2-A^2}=\sqrt{(C-A)(C+A)},
\end{equation}
\begin{equation}\label{M_42}
    C-A=1+a^2r^4,\,\,C+A=1+2r^2+a^2r^4.
\end{equation}
Again,
\begin{equation}\label{M-43}
    \langle K^\prime f_*,f_*\rangle= \int_0^\infty
    r^5\exp\left(-\dfrac12h^2r^2\right)I(r)\,dr
\end{equation}
where
\begin{equation}\label{M-44}
    I(r)=\dfrac1{2\pi} \int_0^{2\pi} \dfrac{(\sin2\varphi)^2}
    {C^2-A^2(\sin2\varphi)^2}\,d\varphi.
\end{equation}
We are lucky to have an explicit integration after observation that
\begin{equation}\label{M-45}
    \int_0^{2\pi}\ldots= 8\int_0^{\pi/4}\ldots
\end{equation}
and substitution $t=\tan2\varphi$. Then
\begin{equation}\label{M-46}
    (\sin2\varphi)^2=\dfrac{t^2}{1+t^2},\,\,
    (\cos2\varphi)^2=\dfrac1{1+t^2}
\end{equation}
\begin{equation}\label{M-47}
    dt=\dfrac{2d\varphi}{(\cos2\varphi)^2}
    \quad
    \text{and}
    \quad
    d\varphi=\dfrac12\cdot\dfrac{dt}{1+t^2}
\end{equation}
and
\begin{multline}\label{M-48}
    I(r)=\dfrac1{2\pi}\cdot8\cdot\dfrac12
    \int_0^\infty\left(-\dfrac1{1+t^2}\right)
    \dfrac{dt}{C^2(1+t^2)-A^2t^2}=\displaybreak[2]\\
    =\dfrac2\pi \int_0^\infty\left(1- \dfrac1{1+t^2}\right)
    \dfrac{dt}{C^2+B^2t^2}.
\end{multline}
By the elementary calculus
\begin{equation}\label{M-49}
    \int_0^\infty \dfrac{dt}{p^2+q^2t^2}=\dfrac\pi2\cdot\dfrac1{pq}, \text{
    any }p,q>0,
\end{equation}
and
\begin{equation}\label{M-50}
    \dfrac1{1+t^2}\cdot\dfrac1{C^2+B^2t^2}=\left(
    \dfrac1{1+t^2}-\dfrac{B^2}{C^2+B^2t^2}\right)\cdot \dfrac1{A^2}.
\end{equation}

Therefore, by \eqref{M-48}--\eqref{M-50} and \eqref{M-40}
\begin{equation}\label{M-51}
    I(r)=\dfrac1{CB}-\dfrac1{A^2}\left(1-\dfrac{B^2}{CB}\right)
    =\dfrac1{A^2}\left[\dfrac CB-1\right]\equiv I_1(r)-I_2(r).
\end{equation}
According to \eqref{M-43} put
\begin{equation}\label{M-52}
    J_\delta=\int_0^\infty r^5\exp\left(
    -\dfrac12h^2r^2\right)I_\delta(r)\,dr,\,\,\delta=1,2.
\end{equation}
Then an easy part gives with $A^2=r^4$
\begin{equation}\label{M-53}
    J_2=\int_0^\infty r\exp\left(-\dfrac12h^2r^2\right)\,dr=
    H^2\int_0^\infty\exp\left(-\dfrac12h^2r^2\right)\,
    d\left(\dfrac12h^2r^2\right)=H^2.
\end{equation}
But with \eqref{M-41}, \eqref{M_42}
\begin{equation}\label{M-54}
    J_1=\int_0^\infty r\exp\left(-\dfrac12h^2r^2\right)
    \dfrac{(1+r^2+a^2r^4)\,dr}{\sqrt{1+a^2r^4}\cdot
    \sqrt{1+2r^2+a^2r^4}}
\end{equation}
and with
\begin{equation}\label{M-55}
    R=2r^2,\,\, r\,dr=\dfrac14\,dR
\end{equation}
\begin{equation}\label{M-56}
    J_1=\int_0^\infty\exp\left( -\dfrac14h^2R\right)
    \dfrac{(1+\frac12R+\frac14a^2R^2)\frac14\,dR}
    {(1+\frac14a^2R^2)^{1/2}(1+R+\frac14a^2R^2)^{1/2}}.
\end{equation}
Finally, put
\begin{equation}\label{M-57}
    w=\frac14h^2R,\,\,R=4H^2w,
\end{equation}
so
\begin{equation}\label{M-58}
    J_1=\int_0^\infty e^{-w}\dfrac{(1+2H^2w+4a^2H^4w^2)H^2\,dw}
    {(1+4a^2H^4w^2)^{1/2}[1+4H^2w+4a^2H^4w^2]^{1/2}}.
\end{equation}
By \eqref{M-53} and \eqref{M-36}, \eqref{M-33} --- an easy part ---
\begin{equation}\label{M-59}
    \dfrac{J_2}{\|f_*\|^2}= \dfrac{H^2}{\sqrt\pi/2H^3}
    =\dfrac2{\sqrt{\pi}}h,
\end{equation}
but
\begin{multline}\label{M-60}
    \widetilde{J}_1=\dfrac{J_1}{\|f_*\|^2}=\displaybreak[2]\\
    =\dfrac2{\sqrt{\pi}}\int_0^\infty e^{-w} \dfrac{H^2\cdot
    2H^2(\frac12h^2+w+2(aH)^2w^2)\,dw}
    {H^3\cdot2H(\frac14h^2+w+(aH)^2w^2)^{1/2}(1+4(aH^2)^2+w^2)^{1/2}}=
    \displaybreak[2]\\
    =\dfrac2{\sqrt{\pi}}\int_0^\infty e^{-w}(1+4(aH^2)^2w^2)^{-1/2}
    \cdot\dfrac{w+\frac12h^2+2(aH)^2w^2}
    {(w+\frac14h^2+(aH)^2w^2)^{1/2}}\,dw.
\end{multline}
Put
\begin{equation}\label{M-61}
    u=4a^2H^4w^2,
\end{equation}
and
\begin{equation}\label{M-62}
    s=\dfrac14h^2+(aH)^2w^2.
\end{equation}
Then
\begin{equation}\label{M-63}
    1-(1+u)^{-1/2}= \dfrac{u}{1+u+\sqrt{1+u}}= \dfrac u2 \cdot
    g_a(w)
\end{equation}
and
\begin{equation}\label{M-64}
    0\leq g_a(w)<1,
\end{equation}
\begin{equation}\label{M-65}
    g(u)\to1\,\text{ if }u\to0;
    \to0\,\text{ if }u\to\infty.
\end{equation}
For the second factor in the integrant on the right side of
\eqref{M-60} we use elementary identities
\begin{multline}\label{M-66}
    \dfrac{w+2s}{\sqrt{w+s}}-w^{1/2}= \dfrac{w+2s-
     (w(w+s))^{1/2}}{\sqrt{w+2s}}=\displaybreak[2]\\
    =\dfrac{s(3w+4s)}{(\sqrt{w+s}(w+2s+w(w+s))^{1/2})} \equiv s\cdot
    G(w)
\end{multline}
where
\begin{equation}\label{M-67}
    G(w)\leq \dfrac{4(w+s)}{(w+s)[\sqrt{w+2s}+\sqrt{w}]}\leq
    2w^{-1/2},
\end{equation}
and
\begin{equation}\label{M-68}
   \begin{array}{l}
    G(w)\to2w^{-1/2}\,\text{ if }\,h\to0,aH\to0,\\
    \text{ or } \frac12w^{1/2}G(w)\to1,\,\text{ if }\,h\to0,aH\to0.
    \end{array}
\end{equation}
After these notations and observations we can continue to evaluate
\eqref{M-60} and write
\begin{multline}\label{M-69}
    \widetilde{J}_1=\dfrac{J_1}{\|f_*\|^2}=
    \dfrac2{\sqrt{\pi}}\int_0^\infty e^{-w}\left(1-\dfrac12ug(w)\right)
    \left[w^{1/2}+sG(w)\right]\,dw=\displaybreak[2]\\
    \dfrac2{\sqrt{\pi}}\int_0^\infty e^{-w}\left[w^{1/2}-\dfrac12
    uw^{1/2}g(w)+sG(w)-\dfrac12usg(w)G(w)\right]\,dw.
\end{multline}
Put correspondingly
\begin{equation}\label{M-70}
    \widetilde{J}_1=F_0-F_1+F_2-F_3
\end{equation}
and let us do calculations and estimates under the assumption
\begin{equation}\label{M-71}
    a\to0,\,h\to0, \,aH^2\to0,\,hH=1.
\end{equation}
Then
\begin{equation}\label{M-72}
    F_0=\dfrac2{\sqrt{\pi}}\int_0^\infty w^{1/2}e^{-w}\,dw=
    \dfrac2{\sqrt{\pi}}\Gamma\left(\dfrac32\right)=1.
\end{equation}
\begin{multline}\label{M-73}
    F_1=\dfrac2{\sqrt{\pi}}\int_0^\infty e^{-w}\dfrac12(4a^2H^4w^2)
     w^{1/2}g_a(w)\,dw=\displaybreak[2]\\
    =\dfrac4{\sqrt{\pi}}a^2H^4 \int_0^\infty
    e^{-w}w^{5/2}g_a(w)\,dw=
    \dfrac{15}2a^2H^4(1+\varepsilon_1(a))
\end{multline}
where $\varepsilon_1(a)\to0$ under conditions \eqref{M-71} by
\eqref{M-61}--\eqref{M-64} and Lebesgue dominated convergence
theorem (LDCT). We use also that
\[
 \Gamma\left(\dfrac72\right)=\dfrac52\cdot\dfrac32\cdot\dfrac12
 \cdot\Gamma\left(\dfrac12\right)=\dfrac{15}8\sqrt\pi.
\]
Next step is an evaluation of $F_2$. By \eqref{M-62} and
\eqref{M-66}, \eqref{M-67}, \eqref{M-68}
\begin{multline}\label{M-74}
    F_2=\dfrac2{\sqrt{\pi}}\int_0^\infty
     e^{-w}sG(w)\,dw=\displaybreak[2]\\
    =\dfrac2{\sqrt\pi}\int_0^\infty e^{-w}\left[\dfrac14
    h^2+(aH)^2w^2\right]G(w)\,dw=\displaybreak[2]\\
    =\dfrac2{\sqrt\pi}\cdot\dfrac14h^2\int_0^\infty e^{-w}G(w)\,dw+
    \dfrac2{\sqrt\pi}(aH)^2\int_0^\infty e^{-w}w^2G(w)\,dw=\displaybreak[2]\\
    =\dfrac1{2\sqrt\pi}h^2\Gamma\left(\dfrac12\right)(1+
    \varepsilon_2(a))+\dfrac2{\sqrt\pi}(aH)^22\Gamma\left(\dfrac52
    \right)(1+\varepsilon_2^\prime)=\displaybreak[2]\\
    =h^2+3(aH)^2+h^2\varepsilon_2(a)+(aH)^2\cdot3
    \varepsilon_2^\prime(a)
\end{multline}
where $\varepsilon_2(a)\to0$, $\varepsilon_2^\prime(a)\to0$ if
\eqref{M-71} holds. One more step evaluates $F_3$.
\begin{multline}\label{M-75}
    F_3=\dfrac2{\sqrt\pi}\int_0^\infty \dfrac12e^{-w}usg(w)G(w)\,dw=
     \displaybreak[2]\\
    =\dfrac1{\sqrt\pi}\int_0^\infty e^{-w}(4a^2H^4w^2)
    \left(\dfrac14h^2+(aH)^2w^2\right)\cdot
    g(w)G(w)\,dw=\displaybreak[2]\\
    =\dfrac1{\sqrt\pi}\int_0^\infty e^{-w}[a^2H^2w^2+4a^4H^6w^4]
    2w^{-1/2}\gamma(w)\,dw
\end{multline}
where $0<\gamma(w)<1$, $\gamma_a(w)\to1$ if \eqref{M-71} holds.
Therefore, by LDCT
\begin{multline}\label{M-76}
    F_3=\dfrac2{\sqrt\pi}a^2H^2\Gamma\left(\dfrac52\right)
    (1+\varepsilon_2(a))+\dfrac8{\sqrt\pi}a^4H^6\Gamma\left(
    \dfrac92\right)(1+\varepsilon_3^\prime(a))=\displaybreak[2]\\
    =\left[\dfrac32a^2H^2+ \dfrac{105}2a^4H^6\right](1+
    \varepsilon_3^{\prime\prime}(a))
\end{multline}
where $\varepsilon_3(a)$, $\varepsilon_3^\prime$,
$\varepsilon_3^{\prime\prime}\to0$ if \eqref{M-71} holds.

Now we need to collect five terms $F_j$, $0\leq j\leq3$, and $J_2$.
Remind orders of these terms
\[
 \begin{split}
  F_0 &{}=1 \text{ by }\eqref{M-72}\\
  F_1 &{}\sim a^2 H^4 \text{ by } \eqref{M-73}\\
  F_2 &{}\sim h^2+(aH)^2 \text{ by } \eqref{M-74} \\
  F_3 &{}\sim a^2H^2+a^4H^6 \text{ by } \eqref{M-76} \\
  \widetilde{J}_2 &{}\sim h\text{ by }\eqref{M-59}
\end{split}
\]
Under conditions \eqref{M-71} with
\begin{equation}\label{M-77}
    a^2H^2=(a^2H^4)\cdot h^2,\quad a^4H^6=(a^2H^4)^2\cdot h^2
\end{equation}
if $F_1\sim\widetilde{J}_2$ two terms $F_1$, $\widetilde{J}_2$
majorize $F_2$, $F_3$, i.e.,
\begin{equation}\label{M-78}
    \dfrac{(F_2+F_3)}h=O(h),
\end{equation}
and
\begin{equation}\label{M-79}
    n(a)=\dfrac{\langle K^\prime f_*,f_*\rangle} {\langle
    f_*,f_*\rangle}=1-\widetilde{J}_2-F_1+O(h^2)=1-
    \dfrac2{\sqrt\pi}h-\dfrac{15}2a^2H^4+O(h^2)
\end{equation}
if
\begin{equation}\label{M-80}
    h\sim a^2H^4\quad\text{or}\quad
    h=\lambda a^{2/5},
\end{equation}
so
\begin{equation}\label{M-81}
    n(a)=1-\left( \dfrac2{\sqrt\pi}\lambda+
    \dfrac{15}2\cdot\dfrac1{\lambda^4}\right)a^{2/5}+O(a^{4/5}).
\end{equation}

We will not get sharp constants in the final result (see
Theorem~\ref{M-teo-8}) but at least at this step we'll anyway choose
the best possible $g_*$ by finding
\begin{equation}\label{M-82}
    g_*=\min\left(\dfrac2{\sqrt\pi}\lambda+\dfrac{15}2\cdot
    \dfrac1{\lambda^4}\right).
\end{equation}
It is attained for $\lambda_*$,
\begin{equation}\label{M-83}
    \lambda_*=(15\sqrt\pi)^{1/5}
\end{equation}
so
\begin{equation}\label{M-84}
    g_*=\dfrac12(5^6\cdot3\cdot\pi^{-2})^{1/5}\approx
    2.718305798\ldots<3.
\end{equation}

Now, in \eqref{M-33} we are ready to choose
\begin{equation}\label{M-85}
    h=\lambda_*a^{2/5}.
\end{equation}
Therefore, we've shown, by \eqref{M-81} and \eqref{M-84}, that
\begin{equation}\label{M-86}
    \dfrac{\langle Kf_*,f_*\rangle} {\langle f_*,f_*\rangle}=
    1-g_*a^{2/5}+O(a^{4/5})\geq1-3a^{2/5}
\end{equation}
for a small enough $a$. We proved \eqref{M-35} and
Proposition~\ref{prop-M-2}.
\end{proof}

3. As we noticed in \eqref{M-21}--\eqref{M-24}
\begin{equation}\label{M-87}
    E^o(a)=\|K^o_{a1}\|.
\end{equation}
To evaluate this quantity from above, we'll use the following Schur
lemma [Sc].
\begin{lemma}\label{lem-3}
Let $A$ be an integral symmetric operator, i.e., $A(x,y)=A(y,x)$,
\[
 A:\,f\to\int_{\mathbb{R}^m}A(x,y)f(y)d\mu
\]
in $L^2(\mathbb{R}^m)$. Then
\[
 \|A\|\leq\sup_x\int_{\mathbb{R}^m}|A(x,y)|d\mu(y).
\]
\end{lemma}
(See more about Schur lemma, or Schur test, in \cite{DK}, Section 3,
or \cite{HS}, Theorem 5.2. More general statements in the context of
the operator interpolation theory can be found in \cite{Mi,Ca}.)

It is quite surprising that this lemma gives us the sharp up to the
second term estimate of the norm $\|K^o\|$.
\begin{prop}\label{prop-4}
Let $K^\prime$ be a kernel \eqref{M-30}. Then
\begin{equation}\label{23.1}
    \sup_x\int_{-\infty}^\infty|K^\prime(x,y)|dy\leq1-\beta(a)
\end{equation}
where
\begin{equation}\label{23.2}
    \beta(a)\geq c\,a^{2/5} \text{ for small enough }a>0,
\end{equation}
$c$ being an absolute constant, $c>2/3$.
\end{prop}
\begin{proof}[Proof \em\ is straightforward]
\begin{equation}\label{23.3}
    K^\prime(x,y)=-K(-x,y)
\end{equation}
so we can consider only $x>0$ in \eqref{23.1}.

Then
\begin{multline}\label{23.4}
    J\equiv\int_{\mathbb{R}}|K^\prime(x,y)|dy=\\
    =\dfrac{2x}\pi\cdot2\int_0^\infty
    \dfrac{ydy}{(1+x^2+y^2+a^2(x^2+y^2)^2)^2-4x^2y^2}
\end{multline}
and after substitution $Y=y^2$
\begin{equation}\label{24.1}
    J=\dfrac{2x}\pi\int_0^\infty
    \dfrac{dY}{(1+x^2+Y+a^2(x^2+Y)^2)^2-4x^2Y}.
\end{equation}
Notice that the denominator
\begin{multline}\label{24.2}
    D=a^4(x^2+Y)^4+2(1+x^2+Y)(x^2+Y)^2a^2+1+2(x^2+Y)+(Y-x^2)^2>\\
    >2a^2x^6+(Y-x^2)^2+2(Y-x^2)+1+4x^2=
    (Y-x^2+1)^2+4x^2+2a^2x^6.
\end{multline}
Therefore, with $\xi=Y-x^2+1$, and $\xi=2xt$,
\begin{multline}\label{24.3}
    J<\dfrac{2x}\pi\int_{-x^2+1}^\infty
      \dfrac{d\xi}{\xi^2+4x^2+2a^2x^6}<\\
    <\dfrac{2x}\pi\int_{-x/2}^\infty
      \dfrac{2xdt}{4x^2\left[t^2+1+\frac12a^2x^4\right]}\equiv p(x).
\end{multline}
By \eqref{M-49}
\begin{equation}\label{24.4}
    p(x)=\left(1+\dfrac12a^2x^4\right)^{-1/2}
    \left[\dfrac12+\dfrac1\pi\arctan\dfrac
    x{2\sqrt{1+\frac12a^2x^4}}\right].
\end{equation}
In \eqref{24.4} we have two positive factors, each of them less than
1, so if we expect their product to be close to 1 we want each of
them to be close to 1. It will be achieved if
\begin{equation}\label{25.1}
    a^2x_*^4\to0,\,\,x_*\to\infty
\end{equation}
when $a\to0$ where
\begin{equation}\label{25.2}
    x_*=x_*(a),\,\,p(x_*)=\max_{x>0}p(x).
\end{equation}
So far we used rough inequalities \eqref{24.2}--\eqref{24.3} and we
do not expect to get sharp constants. Therefore, we do not look for
finding exact $x_*$ but we want reasonable estimates for
\begin{equation}\label{25.3}
    p^*=p(x_*)=\max_{x>0}p(x).
\end{equation}
Notice that for $v>1$
\begin{multline}\label{25.4}
    \dfrac1\pi\arctan v=\dfrac12-\dfrac1\pi\int_v^\infty
      \dfrac{dy}{1+y^2}\leq\\
    \leq\dfrac12-\dfrac1\pi\dfrac{v^2}{1+v^2}
    \int_v^\infty\dfrac{dy}{y^2}<
    \dfrac12-\dfrac1\pi\cdot\dfrac1v\left(1-\dfrac1{v^2}\right)\\
    <\dfrac12-\dfrac7{22}\cdot\dfrac1v\text{ if }v>30.
\end{multline}
Therefore, by \eqref{24.3}, \eqref{25.3}, \eqref{25.4}
\begin{equation}\label{26.2}
    p(x)\leq\left(1+\dfrac12a^2x^4\right)^{-1/2}\left(1-\dfrac7{11}
    \cdot\dfrac1x\right)
\end{equation}
and by elementary inequality
\begin{equation}\label{26.3}
    (1+w)^{-1/2}<1-\dfrac9{20}w\quad\text{if }0<w<\dfrac7{50}
\end{equation}
so with $x>1$
\begin{multline}\label{26.4}
    p(x)<\left(1-\dfrac9{40}a^2x^4\right)\left(1-\dfrac7{11}\cdot\dfrac1x\right)=\\
    =1-\left(\dfrac9{40}a^2x^4+\dfrac7{11}\cdot\dfrac1x\right)+
    \dfrac{63}{440}\cdot a^2x^3<\\
    <1-\left(\dfrac9{110}a^2x^4+\dfrac7{11}\cdot\dfrac1x\right).
\end{multline}
Again, as in \eqref{M-82}--\eqref{M-85} we look for
\begin{equation}\label{26.5}
    g^*=\min_{\mu>0}\left(\dfrac7{11}\cdot\dfrac1\mu+\dfrac9{110}\mu^4\right)
\end{equation}
and
\begin{equation}\label{26.6}
    \mu^*=\left(\dfrac7{11}\cdot\dfrac{110}{4\cdot9}\right)^{1/5}=
    \left(\dfrac{35}{18}\right)^{1/5}\approx1.142244585.
\end{equation}
The choice
\begin{equation}\label{27.1}
    x^*=\mu^*a^{-2/5}
\end{equation}
gives the best result in the right side of \eqref{26.4}. This leads
us to the inequality
\begin{equation}\label{27.2}
    p(x)<1-g^*a^{2/5}
\end{equation}
where by \eqref{26.5}
$g^*=\dfrac{35}{44}\left(\dfrac{18}{35}\right)^{1/5}\approx
0.696395987$
\begin{equation}\label{27.3}
    g^*>\dfrac23.
\end{equation}
We proved \eqref{23.1}--\eqref{23.2} with $c=\dfrac23$.
\end{proof}

4. In \eqref{M-33} we've chosen a smooth cut-off but calculations of
Sect.~2 could be done (as long as we do not try to find sharp
constants) with other $f_*$'s, say,
\begin{equation}\label{28.1}
    f_*(x)=\left\{
     \begin{array}{ll}
      x, & |x|\leq H,\\
      0, & |x|>H.
     \end{array}
    \right.
\end{equation}
Then
\begin{equation}\label{28.2}
    \langle f_*,f_*\rangle=\dfrac23H^3.
\end{equation}
Again, the integral \eqref{M-43} will play important role, i.e., we
use the following.
\begin{lemma}\label{lem-5}
If $C>A>0$ then
\begin{equation}\label{28.3}
    I=\dfrac1{2\pi}\int_0^{2\pi} \dfrac{(\sin2\varphi)^2d\varphi}
    {C^2-A^2(\sin2\varphi)^2}= \dfrac1{A^2}\left[\dfrac CB-1\right]
\end{equation}
where
\begin{equation}\label{28.4}
    A^2+B^2=C^2,\,\,B>0.
\end{equation}
\end{lemma}
\begin{proof}[Proof \em was given in Sect~2, formulas
\eqref{M-44}--\eqref{M-51}]

Now
\begin{equation}\label{28.5}
    n:=\langle K^\prime f_*,f_*\rangle=\iint_{|x|,|y|\leq H}
    K^\prime(x,y)xydxdy
\end{equation}
and with the integrand being positive we have
\begin{equation}\label{29.1}
    \int_0^H\left[\dfrac CB-1\right]rdr\leq n\leq \int_0^{H\sqrt2}
    \left[\dfrac CB-1\right]rdr.
\end{equation}
The same analysis as on pp.~\pageref{M-54}--\pageref{M-77} will
shows that if
\[
 H^{-1}=h=\lambda\,a^{2/5}
\]
then
\[
 \dfrac n{2/3H^3}=1-\varphi(a)
\]
where
\[
\varphi(a)\leq C_4a^{2/5}
\]
although because of \eqref{29.1} with different upper bounds of
integration this absolute constant $C_4$ will be worse than in
\eqref{M-84} or \eqref{M-86}, even if we will try to choose
$\lambda$ appropriately.
\end{proof}

5. In previous section we saw that Schur lemma gives good upper
estimates \eqref{23.1}--\eqref{23.2} 
 of the norm of an
integral operator with the kernel $K^\prime\in$\eqref{M-30},
\eqref{M-32}. But an attempt to apply Lemma~\ref{lem-3} to the
kernel $K\in\eqref{M-2}$, $T=1$, does not give a right order of the
term which is an analogue of $\beta$ in \eqref{23.1}. Even if we
take $x=0$
\begin{multline*}
    \int_{\mathbb{R}}K(0,y)\,dy=
    \dfrac2\pi\int_0^\infty\dfrac{dy}{1+y^2+a^2y^4}=\displaybreak[2]\\
    =1-\dfrac2\pi a^2\int_0^\infty\dfrac{y^4\,dy}
    {(1+y^2)(1+y^2+a^2y^4)}=\displaybreak[2]\\
    =1-C(a)\cdot a\quad\text{with}\quad
    C(a)\to1\,\,(a\to0).
\end{multline*}
So even if the estimate
\[
 \|K_a\|\leq1-\dfrac34a,\quad a\leq a^*
 \quad\text{for some}\quad a^*>0
\]
were correct it would be far away from the below estimate
\eqref{M-86}.

However, a more skillful use of Schur lemma (or its proof) combined
with Uncertainty Principle (in its additive form) gives (!) good
estimates of the norm $\|K_a\|$ of the full operator \eqref{M-1},
\eqref{M-2}. These constructions have been suggested by Fedor
Nazarov [private communication, Oct. 26, 2006].
\begin{lemma}[Uncertainty Principle]\label{lem-6}
Let $f\in L^2(\mathbb{R}^1)$,
$\displaystyle\int_{\mathbb{R}^1}|f(x)|^2\, dx=1$. For any $h>0$,
$hH=1$, one of two inequalities (a) or (b) holds:

(a)~$\displaystyle\int_{|x|\geq H}|f(x)|^2\,dx\geq\dfrac19$, or

(b)~$\displaystyle\int_{|s|\geq
h}|\widetilde{f}(s)|^2\,ds\geq\dfrac19$,

\noindent where
\[
 \widetilde{f}(s)=\mathcal Ff\equiv\dfrac1{\sqrt{2\pi}}
 \int_{-\infty}^\infty e^{-isx}f(x)\,dx,
\]
i.e., $\mathcal F$ is standard unitary Fourier operator in $L^2$.
\end{lemma}
This is a version of the celebrated
Landau--Pollack--Slepian 
inequalities (Additive Uncertainty Principle). In Appendix we'll
discuss it and give a proof of Lemma~\ref{lem-6} to make the present
paper a self-contained exposition.

Now we will give an estimate from above of the norm of $K$-image,
$K\in\eqref{M-1},\eqref{M-2}$, $T=1$,
\[
 Kf(x)=\int_{-\infty}^\infty K(x,y)f(y)\,dy.
\]

We can assume [see \eqref{M-18}--\eqref{M-25}] that $f(x)\geq0$ if
\[
 \|K\|=\|Kf\|,\,\,\|f\|=1.
\]
If in Lemma~\ref{lem-6} (b) holds we do the following estimates:
\[
 Kf(x)\leq K_0f(x)=\dfrac1\pi \int_{-\infty}^\infty
 \dfrac{f(y)\,dy}{1+(x-y)^2}=k*f,
\]
and $\mathcal F$ being isometry
\begin{multline}\label{M-4.3}
    \|Kf(x)\|_2^2\leq \|k*f\|_2^2=
    \|\widetilde{k}(s)\cdot\widetilde{f}(s)\|_2^2=\displaybreak[2]\\
    =\int_{-\infty}^\infty e^{-2|s|}|\widetilde{f}(s)|^2\,ds=
    1-\int_{-\infty}^\infty\left(1-e^{-2|s|}\right)
    |\widetilde{f}(s)|^2\,ds\leq\\
    \leq1-\left(1-e^{-2h}\right)\cdot\dfrac19<1-\dfrac16h,\,
    \text{if}\,h\leq\dfrac14.
\end{multline}
We used (b) and elementary inequalities
\[
 1-e^{-v}\geq\dfrac34v
\]
if
\[
 0<v<\dfrac12.
\]
If (a) holds we do as Schur did, i.e., by Cauchy inequality, with
$K\cdot f=K^{1/2}(K^{1/2}f)$,
\[
 (Kf(x))^2\leq\int K(x,\xi)\,d\xi\cdot
 \int K(x,y)f^2(y)\,dy\leq\int K(x,y)f^2(y)\,dy
\]
and
\[
 \int(Kf(x))^2\,dx\leq \int M(y)f^2(y)\,dy
\]
where
\[
 M(y)=\int_{-\infty}^\infty \dfrac1\pi\cdot \dfrac{dx}
 {1+(x-y)^2+a^2(x^2+y^2)^2}.
\]
For any $y$ $M(y)\leq1$ but if $|y|\geq H$ we get a better estimate:
notice that if $|x-y|\leq1$ then
\[
 1+(x-y)^2+a^2(x^2+y^2)^2\geq [1+(x-y)^2]
 \left(1+\dfrac{a^2}2y^4\right),
\]
so
\begin{multline}\label{M-6.1}
    M(y)\leq \int_{|x-y|\leq1}
    \dfrac1\pi\dfrac{dx}{\left(1+(x-y)^2\right)
    \left(1+\dfrac{a^2}2y^4\right)}+
    \int_{|x-y|\geq1}\dfrac1\pi\dfrac{dx}
    {1+(x-y)^2}\leq\displaybreak[2]\\
    \leq\dfrac12\cdot\dfrac1 {1+\frac12a^2H^4}+\dfrac12=
    1-\dfrac12\left(1-\dfrac1{1+\frac12a^2H^4}\right)<
    1-\dfrac16a^2H^4
\end{multline}
if $a^2H^4<1$. Let us choose such $h>0$ that
\[
 \dfrac16h=\dfrac16a^2H^4,\,\, hH=1,
\]
i.e.,
\[
 h=\left(a^2\right)^{1/5}.
\]
Then inequalities \eqref{M-4.3} and \eqref{M-6.1} give the same
estimate for the case (b) and (a), correspondingly. By
Lemma~\ref{lem-6}, it covers all possible cases. These inequalities
give the upper bound of the square of the norm $\|K\|^2$; therefore
\[
 \|K\|\leq\left(1-\dfrac16h\right)^{1/2}<
 1-\dfrac1{12}a^{2/5}.
\]
Hence, we proved the following
\begin{prop}\label{prop-7}
Let $K_a$ be an integral operator in $L^2(\mathbb{R}^1)$ with a
kernel
\begin{equation}\label{M-6.5}
    K_a(x,y)=\dfrac1\pi\cdot
    \dfrac1{1+(x-y)^2+a^2(x^2+y^2)^2}.
\end{equation}
Then
\[
 \|K_a\|=E(u)=1-\gamma(a)
\]
where $\gamma(a)\geq\dfrac1{12}a^{2/5}$ for small enough $a>0$.
\end{prop}

Prop \ref{prop-M-2} gives estimates from above of the deficiency
term $\psi$ in \eqref{M-9} in the case of the subspace of odd
functions. But with inequalities \eqref{M-12} of
Lemma~\ref{lem-M-1}, Prop \ref{prop-7} and \ref{prop-M-2} together
complete the proof of the following statement.
\begin{teo}\label{M-teo-8}
Let $E(a)=E^e(a)$ and $E^o(a)$ be the largest eigenvalues of the
integral operator $K_a\in\eqref{M-6.5}$ on subspaces of even and odd
functions correspondingly. Then for $a>0$ small enough
\[
 3a^{2/5}\geq1-E^o(a)>1-E^e(a)=1-E(a)\geq\dfrac1{12} a^{2/5}.
\]
\end{teo}

6. Now we can give an asymptotic of $\tau(a)$ in the solution
\eqref{T-10} of the equation $E^o(a,T)=1$.
\begin{lemma}\label{lem-9}
The norm $N(a,T)$ of an operator $K_{aT}\in$\eqref{M-1} has the
property
\begin{equation}\label{M-8.2}
    N(ab,T/b)=bN(a,T),\,\,\forall \,T,a,b>0.
\end{equation}

The same is true for the norms $N^e$, $N^o$ of $K_{aT}^e$,
$K_{aT}^o$, restrictions of $K\in\eqref{M-1}$ to the subspaces of
even and odd functions.
\end{lemma}
\begin{proof}
Notice that the kernel $K\in\eqref{M-1}$, with $x=b\xi$, $y=b\eta$,
becomes
\begin{multline*}
 K(b\xi,b\xi)=\dfrac1\pi\cdot
 \dfrac1{T^2+b^2(\xi-\eta)^2+a^2b^4(\xi^2+\eta^2)^2}=\displaybreak[2]\\
 =\dfrac1{b^2}\cdot\dfrac1\pi\cdot
 \dfrac1{(T/b)^2+(\xi-\eta)^2+(ab)^2(\xi^2+\eta^2)^2}.
\end{multline*}
However,
\begin{multline}\label{M-9.1}
    \sup_{\|\varphi\|=1} \iint K(b\xi,b\eta)
    \varphi(\xi)\varphi(\eta)\,d\xi d\eta=\\
    =\sup_{\|\varphi\|=1}\dfrac1b\iint K(x,y)
    \dfrac{\varphi(x/b)}{\sqrt b}\cdot
    {\varphi(y/b)}{\sqrt{b}}\,dxdy= \dfrac1bN(a,T),
\end{multline}
because $\|\psi\|=1$, where $\psi(x)=\dfrac1{\sqrt b}\varphi(x/b)$.
It implies
\begin{equation}\label{M-9.2}
    \dfrac1{b^2}N(ab,T/b)=\dfrac1b N(a,T),
\end{equation}
i.e., \eqref{M-8.2} holds.
\end{proof}

If we take in \eqref{M-9.1} only odd $\varphi$ we come to an
identity
\begin{equation}\label{M-9.3}
    N^o(ab,T/b)=bN^o(a,T)
\end{equation}
as well.

The same comment leads to such an identity for $N^e$ although we do
not need to say this because $N^e\equiv N$ anyway.

Put $b=T$ in \eqref{M-9.3}; then we have
\begin{equation}\label{M-10.1}
    N^o(aT;1)=TN^o(a,T),
\end{equation}
and the equation $E^o(a,T)=1$, or --- the same --- $N^o(a,T)=1$,
becomes an equation
\begin{equation}\label{M-10.3}
    N^o(aT;1)=T.
\end{equation}

By Prop~\ref{prop-M-2} and \ref{prop-4}
\begin{equation}\label{M-10-.4}
    N^o(t;1)=1-\psi(t)
\end{equation}
and
\begin{equation}\label{M-10.5}
    c_1t^{2/5}\leq\psi(t)\leq c_2t^{2/5}
\end{equation}
for small $t$. If $T=1-\tau(a)$ and $\tau(a)\to0$ ($a\to0$) as in
\eqref{T-10}, the equation \eqref{M-10.3} links $\tau$ and $\psi$:
\begin{equation}\label{M-11.1}
    1-\psi(aT)=T
\end{equation}
and
\begin{equation}\label{M-11.2}
    \psi(a(1-\tau(a))=\tau(a).
\end{equation}
With $\tau(a)\to0$
\[
 \frac12<1-\tau(a)<1,
\]
so \eqref{M-11.2} and \eqref{M-10.5} imply
\begin{multline}\label{M-11.3}
    C_1\left(\dfrac12a\right)^{2/5}\leq C_1[a(1-\tau(a))]^{2/5}\leq
    \psi(a(1-\tau(a)))\equiv \tau(a)\leq\\
    \leq C_2[a(1-\tau(a))]^{2/5}\leq C_2a^{2/5}.
\end{multline}
We proved the following.
\begin{prop}\label{prop-10}
The temperature shift $\tau(a)\in$\eqref{T-10} has estimates
\begin{equation}\label{M-11.4}
    ca^{2/5}\leq\tau(a)\leq Ca^{2/5}
\end{equation}
for small enough $a>0$, where $c$, $C$ are absolute constants.
\end{prop}

\textbf{Remark}. If we would know that $\lim\psi(a)a^{-2/5}$ existed
and were equal to $L$, then the same argument would tell us that
\[
 \lim_{a\to0}\tau(a)\cdot a^{-2/5}=L.
\]

7. Comments and questions.

7.1. Proposition \ref{prop-10} and Theorem \ref{M-teo-8} give
two-side estimates for $\varphi(a)$, $\psi(a)$ and $\tau(a)$ --- see
\eqref{M-8}--\eqref{T-10}, but no information about existence of the
limits $L^e$ or $L^o=L$ or their numerical values. (We explained
this would-be equality in Remark above.)

But a natural conjecture would go far beyond these limits.

Let $\{E_j(a)\}$, $E_0(a)\geq E_1(a)\geq\ldots$, be a sequence of
eigenvalues of an operator $K_a$. For any $j=0,1,\ldots$
$E_j(a)=1-\varphi_j(a)$, and --- we would conjecture ---
\[
 \lim\varphi_j(a)a^{-2/5} =\mu_j \,\text{ exists},
\]
and $\{\mu_j\}$ are eigenvalues of a (somewhere hidden)
pseudo-differential operator $M$.

This conjecture is inspired by H.~Widom's analysis
\cite{Wi1,Wi2,Wi3} of integral operators
\[
 T_a:\,f\to \int_{-1}^1\dfrac1a\rho\left( \dfrac{x-y}a\right)f(y)
 \,dy
\]
in $L^2(I)$, $I=[-1;+1]$, \cite{KMS,Pa}. He assumes that
\[
 R(s)=\int_{-\infty}^\infty e^{ixs}\rho(x)\,dx
\]
satisfies the following:
\begin{enumerate}
    \item[(i)] $\lim R(s)=0$ when $s\to\pm\infty$.
    \item[(ii)] $\max\{R(\xi):\,|\xi|>\delta\}<M$ for each
    $\delta>0$.
    \item[(iii)] $\displaystyle\lim_{s\to0} |s|^{-\alpha}
    (M-R(s))=c$ ($0<c<\infty$; $0<\alpha<\infty$).
\end{enumerate}
Then a positive definite kernel $V(x,y)$ is given, and its
eigenvalues $\lambda_1\geq \lambda_2\geq\ldots$ are linked with
eigenvalues $\mu_j(T_A)\geq\mu_{j+1}(T_A)$ in the following way. For
fixed $j$
\[
 \lim_{a\to0}(M-\mu_j(T_A))a^{-\alpha}=c\lambda_j ^{-1};
\]
moreover, each sequence of $A$'s ($Aa=1$) tending to infinity has a
subsequence for which $\psi_j(a)$, $T_A\psi_j(a)=\mu_j\psi_j(a)$,
converges in $L^2(I)$ to an eigenfunction of $V$ belonging to the
eigenvalues $\lambda_j$. See details in \cite{Wi2}.

Using Weyl symbols H. Widom gave (private communication) a heuristic
argument which leads to a conjecture that this operator $M$ exists,
it has a symbol $|s| + 4 x^4,$ or in other terms it is determined by
the quadratic form
$$ <Mf, f> \;=\; <|s| \tilde{f}(s), \tilde{f}(s)> + <4x^4 f(x), f(x)>.$$

7.2. An integral operator \eqref{M-1}--\eqref{M-2} was brought to my
attention by P.~Krotkov and their analysis of models in
superconductivity. From mathematical point of view, the kernel
\eqref{M-2} is interesting because
\begin{enumerate}
    \item[--] it is NOT translation invariant,
    \item[--] a polynomal in the denominator is NOT homogenuous, it
    has terms of order 2 and 4.
\end{enumerate}

Although our analysis and results could be extended to a broader
family of such kernels, the complete understanding of an interplay
of orders of terms depending on $(x-y)$ and $(x+y)$, or $(x^2+y^2)$,
would be very instructive.

Notice, for example, that the following is true.
\begin{prop}\label{prop-11-m}
Let
\begin{equation}\label{M-15.4}
    K_a(x,y)=\dfrac1\pi\cdot \dfrac1 {1+(x-y)^2+a^2(x^2+y^2)^t},
\end{equation}
$t>0$ fixed, $a>0$ goes to zero. Then its norms $N=N^e$ and $N^o$
satisfy inequalities
\begin{equation}\label{M-16.1}
    ca^{2/(1+2t)}\leq 1-N^e(a)<1-N^o(a)\leq Ca^{2/(1+2t)}, \,\,
    a\leq a^*,
\end{equation}
where $c$, $C$ are constants depending on $t$ but not on $a$.
\end{prop}

The operator $K_a$ with a kernel \eqref{M-15.4} is compact for any
$t>0$, $a>0$.

The conjecture of Section~7.1 can easily be formulated for this
example as well. How to prove it?

Acknowledgement. The author had interesting and useful discussions
with F.~Nazarov and H.~Widom; he thanks them for valuable remarks
and insights. The present paper has been completed in Fall 2007
during my stay at Weizmann Institute of Science, Rehovot, Israel, as
Weston Visiting Professor. I thank Weizmann Institute for
hospitality and stimulating environment.

\begin{center}
Appendix\\ Additive Uncertainty Principle.
\end{center}

To make our paper self-contained we'll give a proof of
Lemma~\ref{lem-6}. Some details follow closely to \cite{FS},
Sect.~8.

Let $P$, $R$ be two orthogonal projectors in a (real or complex)
Hilbert space $H$. Put $E=\Im P$, $L=\Im R$.
\begin{lemma}\label{lemm-3-1}
The norm of the product
\begin{equation}\label{3-1}
    \|RP\|=b
\end{equation}
where
\begin{equation}\label{3-2}
    b=\sup\{\langle u,v\rangle\,|\,Pu=u,Qv=v, \|u\|\leq1,
    \|v\|\leq1\}
\end{equation}
and
\begin{equation}\label{3-3}
    b^2=\|PRP\|.
\end{equation}
\end{lemma}

Indeed,
\begin{equation}\label{3-4}
    \|RP\|=\sup_{\|f\|,\|g\|\leq1} \langle RPf,g\rangle
\end{equation}
but by \eqref{3-2}
\begin{equation}\label{3-5}
    \langle RPf,g\rangle=\langle Pf,Rg\rangle= \|Pf\|\cdot
    \|Rg\|\cdot \langle u^*,v^*\rangle\leq b
\end{equation}
where $u^*=\dfrac{Pf}{\|Pf\|}$, $v^*=\dfrac{Rg}{\|Rg\|}$. On another
side, for some sequence $\{u_n,v_n\}\subset E\times L$,
$\|u_n\|,\|v_n\|\leq1$,
\begin{equation}\label{3-6}
    b=\lim\langle u_n,v_n\rangle= \lim\langle Pu_n, Rv_n\rangle\leq
    \|RP\|.
\end{equation}
Therefore \eqref{3-5} and \eqref{3-6} implies \eqref{3-1}.

\eqref{3-3} just means that $\|A\|^2=\|A^*A\|$ where
\begin{equation}\label{3-7}
    A=RP,\quad A^*=PR \quad\text{ and }\quad A^*A=PRP.
\end{equation}

The geometric core of this Appendix is
\begin{lemma}\label{lem-3-2}
Let $u,v$ be unit vectors in $H$, and
\begin{equation}\label{3-8}
    \langle u,v\rangle=T,\,|T|=t.
\end{equation}
Then for any $g\in H$, $\|g\|=1$.
\begin{equation}\label{3-9}
    q^2:=|\langle g,u\rangle|^2+|\langle g,v\rangle|^2\leq 1+t.
\end{equation}
\end{lemma}

{\em Proof}. If $t=0$ this is Pythagor's identity. If $t=1$ this is
Cauchy inequality.

\sloppy In the case $0<t<1$ we can choose a subspace $K$, $\dim
K=3$, $K\supset  \text{LinSpan}\{u,v,g\}$ and an o.~n.~b.\
$(e_j)_1^3$ in $K$ in such a way that
\begin{equation}\label{3-10}
 \left\{
    \begin{array}{l}
     u=(1,0,0)\\
     v=(\overline{T},\tau,0),\tau=\sqrt{1-|T|^2}\geq0\\
     g=(x,y,z)\in\mathbb C^3,|x|^2+|y|^2+|z|^2=1.
    \end{array}\right.
\end{equation}
\fussy

For any $h$, $0<h=1/H$ by Cauchy inequality
\begin{multline}\label{3-11}
    q^2=|x|^2+|xT+y\tau|^2=\displaybreak[2]\\
    =|x|^2(1+t^2)+|y|^2\tau^2+2\Re(HxT) (\overline{y}\tau
    h)\leq\displaybreak[2]\\
    \leq |x|^2(1+t^2+H^2t^2)+ |y|^2(1-t^2)(1+h^2)
\end{multline}
and the choice $h=\dfrac t{1-t}$ gives an inequality \eqref{3-9}
\[
    q^2\leq(|x|^2+|y|^2)(1+t)\leq(1+t).
 \eqno{\square}
\]

We are ready to prove the following
\begin{prop}\label{prop-3-3}
For any $f$, $\|f\|=1$,
\begin{equation}\label{3-13}
    r^2:=\|f-Pf\|^2+\|f-Rf\|^2\geq1-b.
\end{equation}
\end{prop}
\begin{proof}
$P$ is an orthogonal projector so $\langle Pf,f-Pf\rangle=0$; then
\begin{equation}\label{3-14}
    \langle f,Pf\rangle=\|Pf\|^2,\,\, \|Pf\|=\langle f,u\rangle
\end{equation}
where $u=Pf/\|Pf\|$ and
\begin{equation}\label{3-15}
    1=\|f\|^2=\|Pf\|^2+\|f-Pf\|^2.
\end{equation}
The same argument gives a vector
\begin{equation}\label{3-16}
    v=Rf/\|Rf\|
\end{equation}
such that
\begin{equation}\label{3-17}
    \|Rf\|=\langle f,v\rangle
\end{equation}
and
\[
 1=\|Rf\|^2+\|f-Rf\|^2.
\]
Therefore, \eqref{3-13} can be rewritten as
\begin{equation}\label{3-18}
    1-|\langle f,u\rangle|^2+1-|\langle f,v\rangle|^2\geq 1-b
\end{equation}
or
\begin{equation}\label{3-19}
    |\langle f,u\rangle|^2+|\langle f,v\rangle|^2\leq 1+b.
\end{equation}
But
\begin{equation}\label{3-20}
    |\langle u,v\rangle|=|\langle Pu,Rv\rangle| =|\langle
    RPu,v\rangle|\leq b
\end{equation}
by Lemma~\ref{lemm-3-1}. Now \eqref{3-19} follows from \eqref{3-9}
of Lemma~\ref{lem-3-2}.
\end{proof}
\begin{exam}
\em If $V$ is a unitary operator in $H$ then
\begin{equation}\label{3-5.0}
    Q=V^*RV
\end{equation}
is an orthogonal projector as well:
\begin{equation}\label{3-5.1}
    Q^2=V^*RVV^*RV= V^*RRV=V^*RV
\end{equation}
and
\begin{equation}\label{3-5.2}
    Q^*=V^*R^*V=V^*RV=Q.
\end{equation}
\end{exam}

The previous statements for $P$, $Q$ and unitarity of $V$ and $V^*$
give us
\begin{equation}\label{3-5.3}
    b=\|QP\|=\|V^*RVP\|=\|RVP\|
\end{equation}
and
\begin{equation}\label{3-5.4}
    b^2=\|(RVP)^*RVP\|=\|PV^*RVP\|=\lambda_0.
\end{equation}

[If this block $T_0=PTP$ of an operator
\begin{equation}\label{3-5.5}
    T=V^*RV
\end{equation}
is compact then
\begin{equation}\label{3-5.6}
    b^2=\lambda_0,\,\lambda_0=\lambda_0(T)
\end{equation}
being its highest eigenvalue.]

Moreover, By Proposition~\ref{prop-3-3}, for any $f$, $\|f\|=1$,
\begin{equation}\label{3-5.7}
    \|f-Pf\|^2+\|f-Qf\|^2\geq1-b.
\end{equation}
However
\[
 \|f-Qf\|=\|f-V^*RVf\|=\|Vf-RVf\|
\]
and the inequality \eqref{3-14} by \eqref{3-5.1} and \eqref{3-5.5}
can be rewritten as:
\begin{equation}\label{3-6.1}
    \|f-Pf\|^2+\|(Vf)-R(Vf)\|^2\geq1-b.
\end{equation}
We proved the following
\begin{teo}\label{teo-3-5}
Let $P$, $R$ be orthogonal projectors in a (real or complex) Hilbert
space $H$, and $V{:}\,H\to H$ a unitary operator. Then for any $f$,
$\|f\|=1$,
\begin{equation}\label{3-6.2}
    \|f-Pf\|^2+\|Vf-RVf\|^2\geq1-b
\end{equation}
where $1\geq b\geq0$,
\begin{equation}\label{3-6.3}
    b^2=\|PV^*RVP\|.
\end{equation}
\end{teo}

This inequality \eqref{3-6.2} is an Abstract Additive Uncertainty
Principle.
\begin{coral}\label{coral-3-6}
Under conditions of Theorem~\ref{teo-3-5}
\[
 \max\{\|f-Pf\|^2;\,\|Vf-RVf\|^2\}\geq\dfrac{1-b}2.
\]
\end{coral}

Of course, the main example for us is $H=L^2(\mathbb R)$ with a
unitary operator $V=\mathcal{F}$, the Fourier transform
\[
 \widetilde{f}(s)=\mathcal{F} f=\dfrac1{\sqrt{2\pi}} \int_{-\infty}^\infty
 e^{-isx}f(x)\,dx,
\]
and projectors $P=R$ where
\[
 Pf=f(x)\chi_I(x),\quad I=[-1,+1].
\]
In this case
\[
 T_0=PV^*RVP{:}\,f\to\left( \int_{-1}^1\left(\int_{-1}^1 e^{-isx}
 f(x)\,dx\right)e^{-isy}\,ds\right)\chi(x)
\]
or
\[
 \begin{array}{cl}
  \displaystyle
  (T_0f)(y)=\dfrac1\pi\int_{-1}^1\dfrac{\sin(y-x)}{y-x}f(x)\,dx, &
  -1\leq y\leq1,\\
  0, & \text{if }|y|>1.
 \end{array}
\]
Now remaining ``hard analysis'' question is to evaluate the norm
$b^2=\lambda_0$ of this operator. The original paper \cite{SP} gives
the value $0.57258$. We'll give a worse (larger) estimate. It comes
if we use (again!) Schur lemma to claim that
\begin{equation}\label{3-8.1}
    \|T_0\|\leq\dfrac1\pi\max_{|x|\leq1}
    \int_{-1}^1\left|\dfrac{\sin(y-x)}{y-x}\right|\,dy:=B^2.
\end{equation}
A function $h(t)=\dfrac{\sin t}t$ is even, positive on $[-\pi,\pi]$
and monotone decreasing on $[0,\pi]$. Indeed,
\[
 h^\prime(t)=\dfrac{t\cos t-\sin t}{t^2}=-\dfrac1{t^2} \int_0^t
 \tau\sin\tau\,d\tau<0,\,\,0<t\leq\pi.
\]
Therefore,
\begin{multline*}
    \dfrac1\pi\int_{-1}^1h(y-x)\,dy\leq\dfrac2\pi \int_0^1h(t)\,dy<
    \dfrac2\pi\int_0^1\left(1-\dfrac{t^2}6+\dfrac{t^4}{120}\right)\,dt=
    \displaybreak[2]\\
    =\dfrac2\pi\left(1-\dfrac1{18}+\dfrac1{600}\right)<
    .60232
\end{multline*}
and
\[
 b<B<.77608 \quad\text{or}\quad
 \dfrac{1-b}2>.11195>\dfrac19.
\]

This completes the proof of Lemma~\ref{lem-6} in the case $h=H=1$.
General case for $hH=1$ immediately follows if we would use the
following property of Fourier transforms:
\begin{equation}\label{new-162}
 (\mathcal{F} g(rx))(s) = 1/r (\mathcal{F} g)(s/r).
\end{equation}


\end{document}